# Earthquake dynamics constrained from laboratory experiments: new insights from granular materials

Andrea Bizzarri*,[1], Alberto Petri[2] and Andrea Baldassarri[2]

[1] Istituto Nazionale di Geofisica e Vulcanologia, Sezione di Bologna, Italy

[2] CNR, Istituto dei Sistemi Complessi, Dipartimento di Fisica, Università La Sapienza, Roma, Italy

## Abstract

The traction evolution is a fundamental ingredient to model the dynamics of an earthquake rupture which ultimately controls, during the coseismic phase, the energy release, the stress redistribution and the consequent excitation of seismic waves. In the present paper we explore the use of the friction behavior derived from laboratory shear experiments performed on granular materials at low normal stress. We find that the rheological properties emerging from these laboratory experiments can not be described in terms of preexisting governing models already presented in literature; our results indicate that neither rate–and state–dependent friction laws nor nonlinear slip–dependent models, commonly adopted for modeling earthquake ruptures, are able to capture all the features of the experimental data. Then, by exploiting a novel numerical approach, we directly incorporate the laboratory data into a code to simulate the fully dynamic propagation of a 3–D slip failure. We demonstrate that the rheology of the granular material, imposed as fault boundary condition, is dynamically consistent. Indeed, it is able to reproduce the basic features of a crustal earthquake, spontaneously accelerating up to some terminal rupture speed, both sub– and supershear.

## 1. Introduction

The modeling of the physical evolution of a fault system, even in the idealized case of a single, isolated fault and even during the coseismic time window only — where the breakdown processes occur, the stress is released and the seismic waves are excited — actually remains one of the big challenges for modern seismology. This is essentially due to our poor knowledge about the physical and chemical features of the fault system (i.e., its initial conditions) and our epistemic ignorance about a realistic constitutive model able to describe the large number of physical and chemical mechanisms occurring, at different spatial scales, during the coseismic phase. Readers can refer to Bizzarri [2011b] and to section 5 of Bizzarri [2014] — and references cited therein — for thorough reviews on this subject, as well as to Ben–Zion and Sammis [2003] and Rice [2006].

**Andrea Bizzarri et al.**

In order to simulate the evolution of a propagating rupture on a fault system we need a geometrical characterization (namely, the tectonic settings) and the physical characterization (basically the PDEs and the physical laws describing how the accumulated stress is released). The most affordable approach used nowadays is the multi–disciplinary combination, and the mutual interactions, of five pillars: *i*) the theoretical approach, intimately related to rock physics, solid matter physics and microphysical properties, *ii*) the accurate, performing and realistic numerical modeling of faulting processes, *iii*) the analysis of data recorded during or after earthquake sequences, *iv*) the geological observations of faults both in the field and in exhumed samples and finally *v*) the laboratory experiments, performed in conditions ideally comparable (in terms of sliding speed and confining stresses) with those of natural faults [Bizzarri, 2014].

Although these areas separately have progressed and reached even relevant results, unfortunately, a unified vision is still missing, but it is necessary. There is no doubt that in the framework of laboratory experiments several goals have been accomplished in the recent decades. Experiments on intact rocks face up to the challenging problem of measuring forces in a setup without pre–cut sliding interfaces. On the other hand, friction experiments (i.e., the frictional characterization of the behavior of pre–existing faults) have now reached the condition of attaining velocities somehow comparable with those expected during natural ruptures (of the order of several (tens) of m/s). However, normal stresses reached in those laboratory experiments are still dramatically far of being representative of crustal conditions (of the order of several tens of MPa), even in most advanced experimental settings [e.g., Niemeijer et al., 2009; Sone and Shimamoto, 2009]. On the other hand, larger values of confining stress are attained in other configurations (for instance in triaxial tests where samples with pre–cut faults can be stressed with hundreds of MPa or in Griggs rigs where several GPa are common), but the sliding speed is not comparable with that of dynamic events [see, e.g., Stanchits, 2006; Raziperchikolaee, 2021; Soleymania, 2021].

Although natural laboratories and induced seismicity have led to a large knowledge of rupture evolution and frictional processes during slip failure, the effective reproduction of earthquake conditions in laboratory is actually far from being reached and therefore the numerical modeling is an available approach. The introduction of a governing law that describes the chemico–physical mechanisms did come from purely theoretical reasoning [e.g., Andrews, 1976] or from extrapolation to real–world of laboratory experiments in idealized condition [e.g., Ruina, 1983]. Once geophysical evidence made some more elaborated (and perhaps realistic) fault constitutive models available, then modelers promptly implemented them and carefully scrutinized the results [Bizzarri and Cocco, 2006a, 2006b; Rice, 2006; Bizzarri, 2009, 2011a, 2012b; Dunham and Rice, 2008; Rice, 2017 among many others]. A part of the present work falls into this line; experimental data from granular material are employed to simulate the spontaneous propagation of a rupture on a planar fault.

Unfortunately, with the unique exception of Sone and Shimamoto [2009], all recent laboratory results performed at high speed (and with the so–called high velocity rotary shear apparatus, recognized as the most interesting generation of laboratory machines employed to simulate fault processes) were not complemented with a robust mathematical model (or an analytical equation) providing a complete governing model. Data are usually presented by plots and descriptions, but often without a fitting model of the whole dynamic process, which is, on the contrary, one of the essential ingredients to be able to numerically model earthquake failures. Incidentally, we recall here that Spagnuolo et al. [2016] provide a fit of data that expresses only the velocity dependence of the steady state level of friction (see their equation (5)) and this is of feeble utility in the modeling of spontaneous rupture propagation, where friction is expected to be well far from steady state conditions. The same holds for Nielsen et al. [2010], where melting process are considered. Notably, an exception in this context is represented by the model proposed by Chen and Spiers [see equation (5a) in Verbene et al., 2020] which expresses the behavior of frictional resistance as a function of the main dilatancy of the shearing grain pack.

In this study we aim to fulfill two separate goals. By considering results from laboratory experiment performed on granular materials, *i*) we explore whether they can be reconciled with previous laws, already presented in the literature and employed in dynamic models, and *ii*) whether they can be used to simulate the spontaneous propagation of an earthquake rupture developing on a planar fault surface.

One class of friction law widely employed in the synthetic reproduction and in the description of seismic events is the rate– and state–dependent laws. These laws have been initially developed to model the friction between two bare surfaces, but they have been found to have some limited validity for a granular fault [Marone, 1998; Mair, 1999]. Such laws are usually tested and parameterized by abrupt transitions between stationary states, like velocity steps and stop–and–go cycles. On the contrary, during a slip episode, the velocity seldom keeps constant but rather changes in time spanning a wide range of values. This is what happens also in the experiments on a laboratory granular fault that we consider in this work. We therefore try to parameterize the friction laws in this dynamical situation, by directly fitting the friction evolution along the slip.

Although developed and tested in experiments performed at ordinary conditions of pressure and with limited contact surfaces and relative velocities, rate–and state–dependent laws have been widely adopted in the description and simulation of coseismic activity and of other tests on laboratory faults conducted at larger, but still far from geological, pressures. Within the same spirit we adopt the dynamical friction laws derived by low pressure granular fault in the simulations of the earthquake propagation.

To this respect some more considerations can be useful. When a granular medium is subjected to a slow shear generated by a compliant force, nor the shear rate neither the shear stress can be controlled. The system self–organizes into a dynamical state displaying intermittent flow, characterized by slip events with highly fluctuating duration and velocities, separated by sticking phases. In this regime the statistical distribution of velocities, slip durations and extension spans a wide range of values and it is often close to a power law. Such a behavior is the most commonly observed in natural phenomena, and it is shared by many other phenomena that under slow external perturbations display strongly intermittent and fluctuating bursts of activity, often called avalanches. Besides earthquakes, with the celebrated Gutenberg–Richter law [Main, 1996], these include fractures [Petri, 1994], structural phase transitions [Cannelli, 1993] plastic deformation of solids [Dimiduk, 2006; Nicolas, 2018], and many others. This common phenomenology and the displaying of a wide range of scales is attributed to the proximity of a critical point, in its statistical physics acceptation, in the dynamics of the system [Sethna, 2001] and therefore to the possibility of universal and scale–independent dynamics in different systems, characterized by a similar common phenomenology, irrespective of the system [Kawamura, 2012].

Nevertheless, it has been recently observed that such universality can be broken by the onset of new forces in the system [Zadeh, 2019]. Specifically, it has been observed that the slips in sheared granular bed may display different velocity average patterns, depending on slip duration and extension, and that this behavior is connected to the differences in the friction force acting in slips of different size [Baldassarri, 2019] [interestingly, the asymmetry observed in the pattern of larger slips is of the same kind of that observed in earthquakes; Mehta, 2006]. In the following we consider therefore different friction laws, each one corresponding to different classes of slip duration.

The aim is to test if, in view of the universal phenomenology described above, their use, after a suitable rescaling of the parameters, can yield results compatible with real observations and/or with those obtained from the friction laws usually adopted in this frame.

The present paper is organized as follows. In section 2 we present the experimental machine, used to perform the laboratory experiments on granular media, and the experimental data obtained during the simulated stick–slip sequences. Section 3 is devoted to the pivotal question of reconciling actual laboratory data with previously–proposed friction laws and constitutive models known from past and recent scientific literature. In section 4 we introduce the novel numerical method used to simulate spontaneous rupture propagation and in section 5 we present and discuss the results obtained with numerical experiments of truly 3–D, spontaneous ruptures obeying to the laboratory–inferred friction curve. Finally, in section 6 we summarize what we really learn from this study and we highlight some possible intriguing perspectives for the near future.

## 2. Experimental results

### 2.1 Laboratory settings

The laboratory set up is of the kind employed in previous works [e.g., Dalton et al., 2005; Baldassarri et al., 2006; Petri et al., 2008], and includes the most updated version [Baldassarri et al., 2019] of the laboratory machine shown in Figure 1. It consists of an annular channel, having internal and external radii $r_{int}$ = 12.5 cm and $r_{ext}$ = 19.2 cm, respectively and being 12 cm high. The walls and the base of the channel are made of perspex and they are fixed. The channel is filled with a bidisperse mixture of glass beads of diameter 1.5 mm and 2.0 mm in equal weight proportion; the use of such mixture prevents crystallization of the medium and results in a total weight of roughly 9.2 kg. The granular medium can be sheared by a horizontal rotating plate, made of perspex as well. The plate fits the channel so that there is no friction with the walls but it does not allow to the grains to get out. It also has a layer of grains glued at its lower side to enhance the dragging and it is free to move vertically to allow changes in the volume of the underlying medium.

**Andrea Bizzarri et al.**

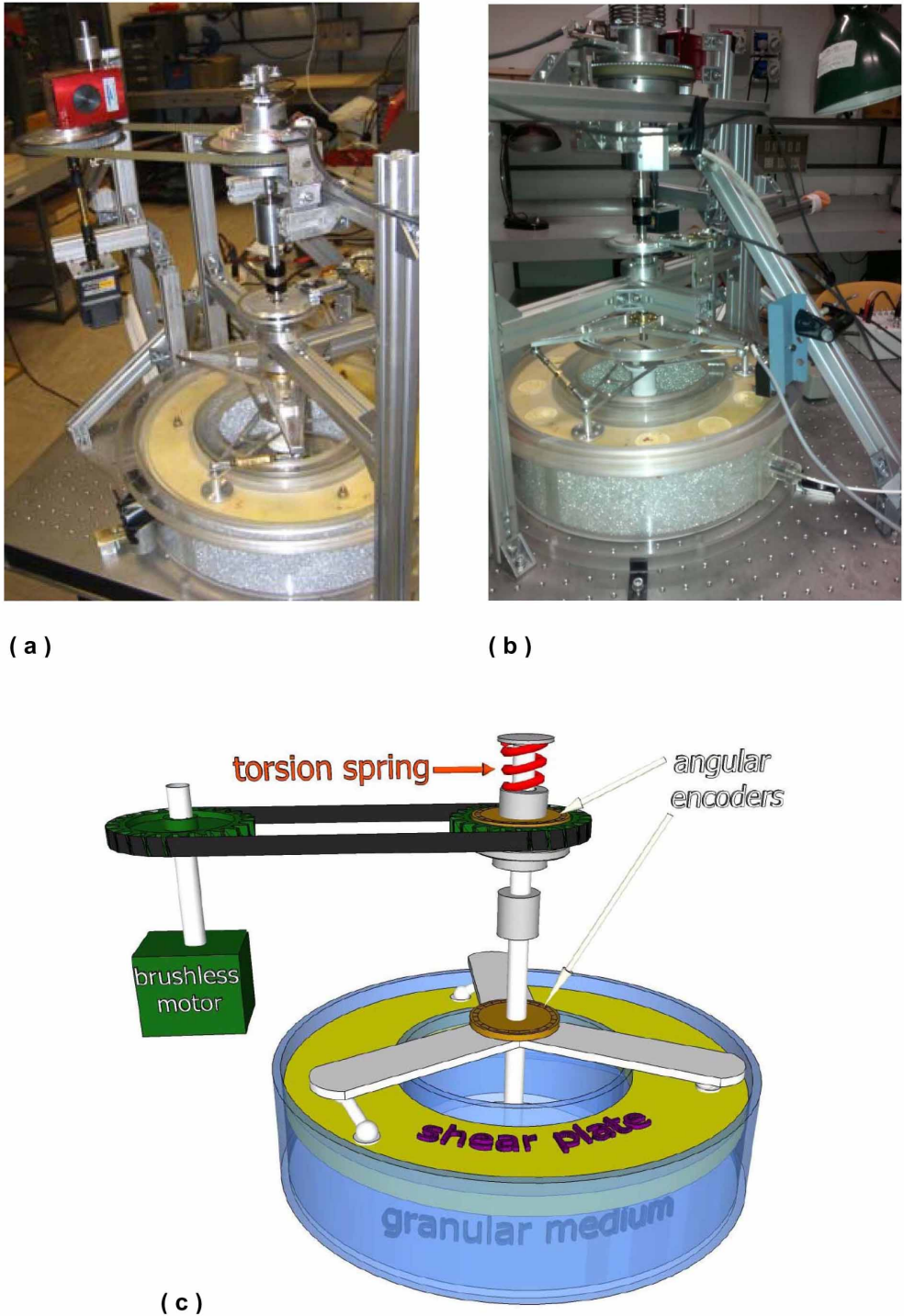

**Figure 1.** Laboratory machine used to perform the experiments on granular materials. (a) and (b) Photo of the installed apparatus. (c) Schematic representation of the machine with the indication of the most prominent parts. Reader can refer to Baldassarri et al. [2006; 2019] for further details.

The mass of the plate being $m$ = 1.2 kg, the medium is subjected to a normal stress $\sigma = \frac{mg}{\pi(r^2_{ext} - r^2_{int})}$ = 176 Pa. This value is kept constant over the whole time duration of each experiment.

Shearing is driven by an electric motor rotating at constant angular speed $\omega_p$ and it can be performed in different ways. For the purpose of the present work the plate has been connected to the motor by a torsion spring of elastic constant $k$ = 0.35 N m. This value has been chosen so that, when driven at low speed, the relative softness of the spring gives rise to an intermittent and irregular — namely *stick–slip* — motion of the plate. For increasing speed the sticks become rarer and the slips longer, until a continuous sliding is observed for driving speeds $\omega_d$ of the order of $\omega_c$ = 0.1 rad/s. This can be perhaps interpreted in term of nucleation length: although a high loading rate after a long stick period would result in the shrinking of the nucleation length, and therefore in an instability [see Guerin–Marthe et al., 2019], a constant high loading rate combined with a stiff system (the spring in our case) would result in weak contacts (or disorganized grains) and therefore in a large nucleation length. This would be therefore more stable; see also the discussion in Rubin and Ampuero [2005].

In order to observe well separated slips we have conducted experiments with the motor driven at an angular velocity $\omega_d$ = 0.015 rad/s, corresponding to about 2.3 mm/s. This appears quite large when compared with speed values usually employed in other laboratory experiments [e.g., Marone, 1998], but consistently with them it is of the order of 1 grain diameter/s. The slip statistics has been shown to be largely independent of $\omega_d$ [Baldassarri et al., 2019] as long as $\omega_d \ll \omega_c$.

The motor axis and the plate axis are supplied with an optical angular encoder having 180,000 divisions, yielding a resolution $\Delta\theta \sim 0.000035$ rad, that is $\Delta u \sim 6$ $\mu$m. The light of a solid state laser reflected by the encoder surface is detected by a photodiode and the resulting signal is acquired at a sampling rate frequency $srf$ = 50 Hz. This allows to know the angular positions of the motor and the plate, $\theta_d$ and $\theta_p$, respectively (and thus their relative position). In addition, one can compute the instantaneous (within 1/50 of second) plate velocity, $\omega_p$, and acceleration. From these quantities the friction torque exerted from the medium, $T$, can be simply computed from the equation of motion: $T = -k(\omega_d t - \theta_p) - I\dot{\omega}_p$, where $I$ is the inertia of the plate. This allows to avoid the insertion of a torque meter in the measurement chain, that we have checked to give the same results and constitutes a further source of noise. Notice that in the case of zero angular acceleration the torque is simply proportional to the spring elongation.

## 2.2 Results from friction experiments on granular materials

In theory, each single slip event (or avalanche) begins when $\omega_p$ starts to differ from zero and it ends when $\omega_p$ goes back to zero. However, in practice it is necessary to choose a threshold value $\omega_{th}$ because of the unavoidable presence of background noise. Therefore, a slip event is identified by the signal portion for which $\omega_p$ remains greater than $\omega_{th}$. The choice of the threshold is to some extent arbitrary; however there is a wide range of small values such that all the observed results are independent of the chosen threshold. For our analysis we have set $\omega_{th} = srf * \Delta\theta \sim 0.00175$ rad/s, which corresponds to a threshold in tangential velocity of about 270 $\mu$m/s. This is the minimum non zero velocity step which can be measured without further smoothing or average of data and corresponds to 1–2 tenths of grain diameter per second. Paradoxically, it increases proportionally to the sampling rate. Along the slip the depth of the shear band changes according to the velocity, with a maximum of about 1cm [Plati, in preparation].

In Figure 2 we report typical results obtained from laboratory experiments. It is seen that, due to the randomness of the granular structure, both velocity (top), and friction (bottom) display very irregular behaviors. However, one can consider their averages, which are quite smooth functions. A recent work [Baldassarri et al., 2019] has discussed their relation with simple models for granular friction and it has shown that in the self–organized regime slips of different duration display very different behavior of the friction. The slips have then been grouped into classes on the basis of their duration $t_{max}$.

As recalled in the introduction, in these kind of experiments $t_{max.}$ is a random variable with a distribution close to a power law [Baldassarri et al., 2019]. Therefore the classes have been chosen equispaced in logarithmic scale. Avalanches at the extremes of the duration distribution have been dropped out when lacking resolution or statistics. Specifically, avalanches shorter than 0.31 s were too small to perform meaningful analysis (less than 15 points at 50 Hz of sampling rate) and those longer than 4.8 s were usually too few. For each class the average slip duration $t_{max}$ and extension $u_{max}$ have been computed (see Table 1). Then the instantaneous values of the plate velocity $\omega_p$ and friction of each slip of the class have been averaged at a set of equispaced discrete times, $\{t_i\}$, such that $0 \leq t_i \leq t_{max}$. In order to smooth out the noise these points are necessarily much less than the number of the original sample points. The number of slip events contributing to each average is also reported in Table 1.

**Andrea Bizzarri et al.**

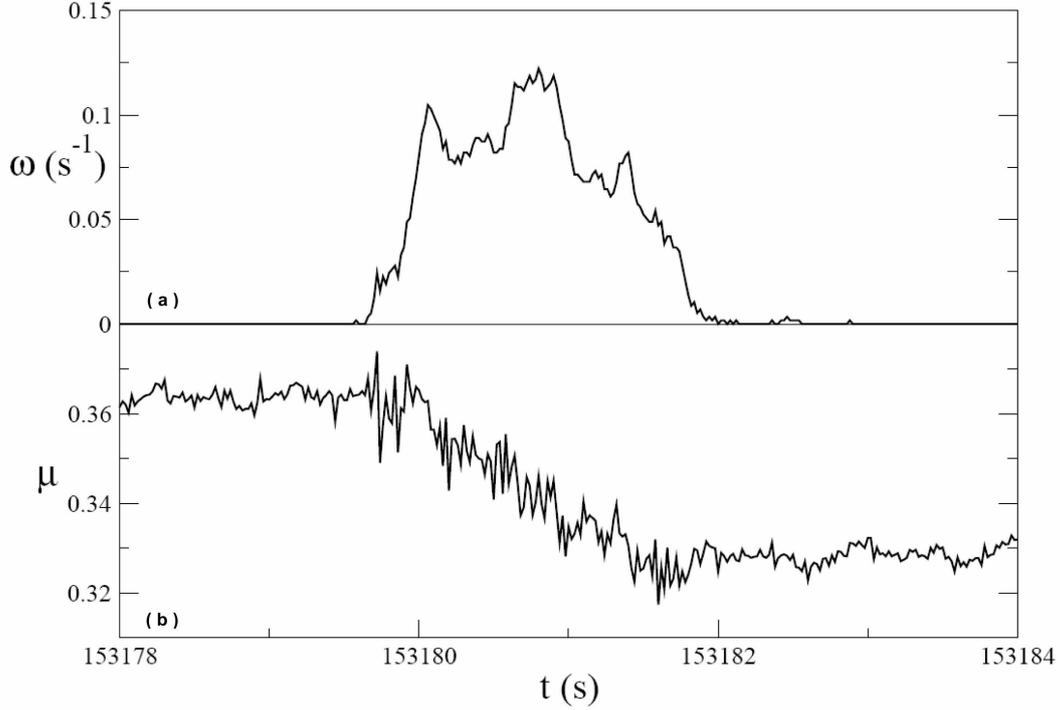

**Figure 2.** Example of the temporal behavior during a single slip. Top panel ((a)) reports the angular velocity and bottom panel ((b)) reports the friction coefficient. The horizontal scale indicates the elapsed time from the beginning of the experiment.

| Experimental dataset (class) | Number of avalanches | $t_{max}$ (s) | $u_{max}$ (rad) | $u_{max}$ (m) |
|---|---|---|---|---|
| 1 | 929 | 0.38 | 0.00302 | 0.00048 |
| 2 | 866 | 0.59 | 0.00687 | 0.00106 |
| 3 | 987 | 0.89 | 0.01495 | 0.00232 |
| 4 | 1694 | 1.34 | 0.03826 | 0.00593 |
| 5 | 1380 | 2.00 | 0.08543 | 0.01324 |
| 6 | 158 | 2.84 | 0.11955 | 0.01853 |

**Table 1.** Different classes of laboratory experiments. For each class we indicate the number of slip events, the average value of the duration of the slip ($t_{max}$) and the average value of the maximum cumulative slip ($u_{max}$). See section 2.2 for further details.

The first two classes exhibit very short slips; being their slip of the order of smaller than the grain size, they display almost constant friction and therefore they are not of interest for our study. The resulting velocity and friction curves for the remaining four classes are shown in Figure 3.

The measured observables are the elapsed time ($t$), the average angular velocity ($\omega_p$) and the average torque ($T$). The angular displacement (Figure 3a) is retrieved from $\omega_p$ exploiting the trapezoidal rule

$$u^{[m+1]} = u^{[m]} + \tfrac{1}{2}\left(\omega_p^{[m+1]} + \omega_p^{[m]}\right)(t^{[m+1]} - t^{[m]}),\, m = 1, \ldots, N-1 \qquad (1)$$

(where $N$ is the number of equispaced points used for averaging of each experiment and $u^{[0]} = 0$), while the apparent friction coefficient $\mu$ is computed from the measured torque $T$ by using the simple equation $\mu = \frac{T}{r}\frac{1}{mg}$, where $r = 0.155$ m is the average radius of the plate (geometric mean of internal and external radii, $r_{int}$ and $r_{ext}$; the arithmetic mean would be 0.159 m) and $g$ the acceleration of gravity.

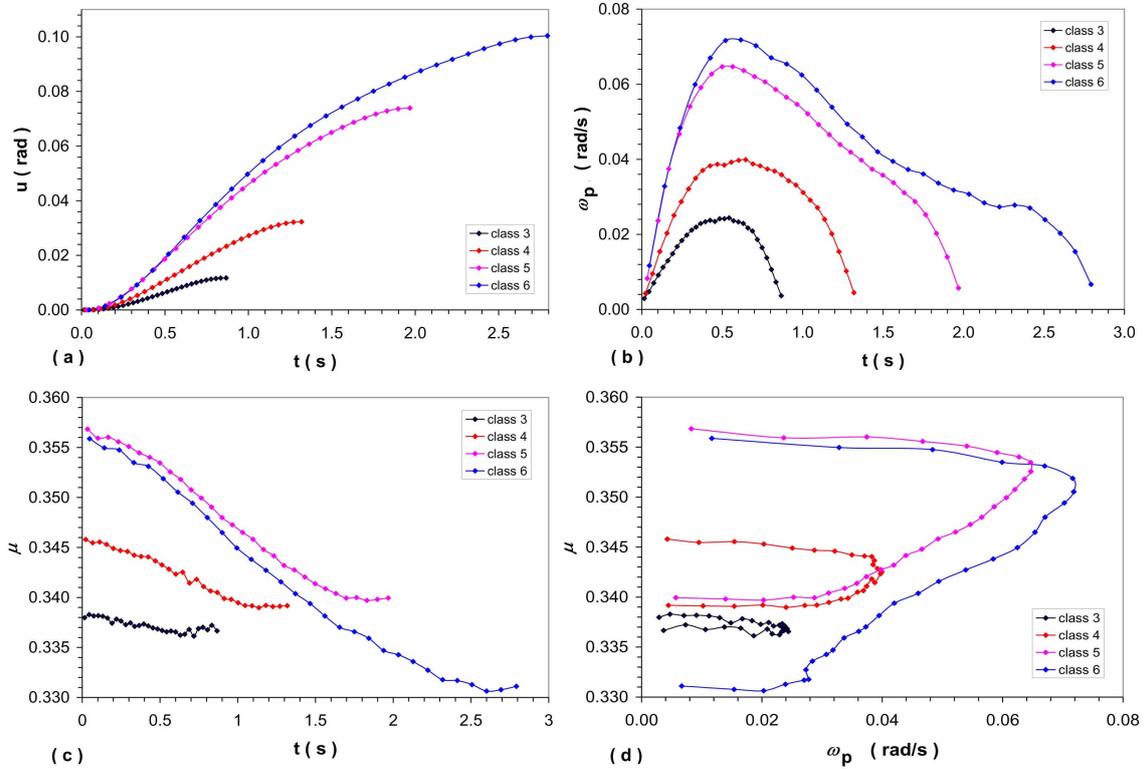

**Figure 3.** Results obtained in laboratory experiments after averaging over many slips of similar duration. The various classes pertain to different values of the duration $t_{max}$ (see Table 1). (a) Time evolution of displacement. (b) Time evolution of angular velocity. (c) Time evolution of the friction coefficient. (d) Phase diagram reporting the evolution of the friction coefficient as a function of the displacement rate. In all panels each square represents an average of the experimental data.

From Figure 3c it is seen that the values of the apparent friction coefficient $\mu$ spans at maximum (class 6) between 0.330 and 0.356. The values are low compared to the typical Byerlee's values $\mu \sim 0.6$–0.8 for rocks [e.g., Byerlee, 1978]. However, these values are frequently encountered in grains and powders [Mair, 1999; GDR MiDi, 2004; Tang et al., 2019, Baldassari et al. 2019], and even in seismic faults and experiments [Collettini, 2019]. They are not proper of the constituent materials, but are due to the granular nature of the medium, as also shown from numerical simulation with materials whose friction coefficient is 0.5 [Mair, 2007]. Comparable values are also found with microscopic models [Van den Ende et al 2018; Ferdowsi and Rubin, 2020]. It can also be observed that there is no stress rise at the beginning of the slip. This is typical not only of the averaged friction, but also of the single slip, as the one in Figure 2. Another example is reported in Figure 1 of Baldassarri [2019], which shows a peak in friction, but well after the slip beginning. As a matter of fact the signal is noisy, with peaks at random positions which average to zero. An initial peak is not observed in many granular systems, even in less noisy simulations [Van den Ende and Niemeijer, 2018].

By looking at Figure 3b it is also apparent that the increasing part of the angular velocity is not abrupt, but it is indeed gradual. Especially for classes 5 and 6 (purple and blue curves, respectively) the time behavior of the angular velocity closely resembles the so–called modified Yoffe function (equation (1) of Bizzarri [2012c]). This result is interesting, because the modified Yoffe function is an analytical regularization [Tinti et al., 2005] of the function

**Andrea Bizzarri et al.**

originally obtained by Yoffe [1951] as a steady state solution of a fixed width, propagating pulse in the mode I of rupture (i.e., for opening cracks) and further extended by Broberg [1978, 1999] and Freund [1979] to the mode II (shear) crack propagation [see Figure 1b, red curve of Bizzarri, 2012c]. Equally interesting is that the qualitative behavior of the experimental time history of angular velocity is also compatible with the closed–form analytical solution that Bizzarri [2012a] found by considering a 1–D spring–slider analogue fault system subject to the linear slip–weakening friction law [see equation (15) of Bizzarri, 2012c and Figure 1b, blue curve of the same paper].

From the phase portrait (Figure 3d) we can see that the peak angular displacement rate does not correspond to the minimum friction. As elucidated by Tinti et al. [2004], this feature is known to correspond to subshear ruptures (i.e., events propagating with a rupture speed smaller than the *S* wave speed of the medium in which the fault is embedded).

In Figure 4 we report the behavior of the friction coefficient as a function of the angular and linear displacement for the four classes of Figure 3 (to obtain values of linear displacement (in meters) one has simply to multiply the angular displacement (in radiants) by the average radius *r*). This plot is in fact a slip–weakening diagram. On the other hand, the fault traction is simply obtained by multiplying the value of $\mu$ by the assumed effective normal stress $\sigma_n^{eff}$. In the present paper we do not consider pore fluid migration and therefore $\sigma_n^{eff}$ is in fact the normal stress of tectonic origin. Vertical arrows indicate the (equivalent) slip–weakening distance, i.e., the amount of displacement at which the friction drop is completed. We can see that the weakening is not linear, especially for classes 5 and 6; we will better discuss this issue in the next section. It is also clear that the laboratory experiments are not able to capture the initial hardening phase of the system, i.e., only the weakening stage is reproduced; indeed, there is a short hardening in the single realizations, see Figure 2, but it is subject to large fluctuations from slip to slip and disappears in averages. In all classes we note a feeble final restrengthening, i.e., a short increase of friction at the end of the experiment, resulting in only a couple of the average point set.

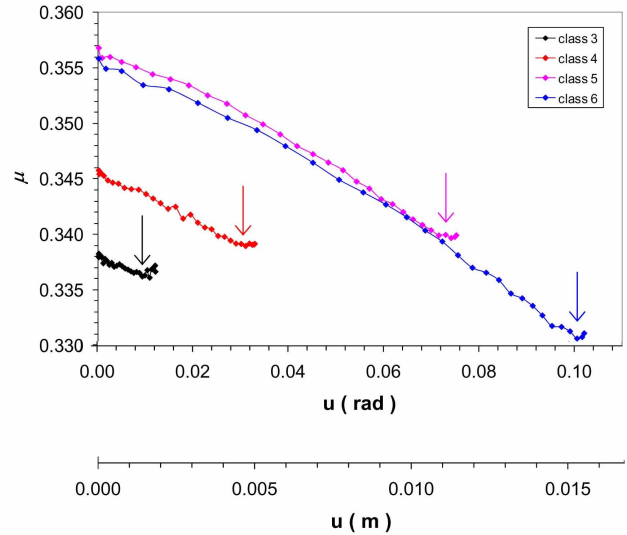

**Figure 4.** Average friction coefficient as a function of angular and linear displacement for the four classes of Figure 3. Angular displacement ($u$(rad)) is computed from equation (1), while linear displacement ($u$(m)) is computed as $u$(rad)*$r$ ($r$ = 0.155 m being the average of internal and external radii of the rotating plate; see section 2). Vertical arrows indicate, for each class, the point at which the stress drop is assumed to be complete. From this value it descends the (equivalent) slip–weakening distance $d_0$.

## 3. Laboratory results and governing models

In this section we will explore whether the experimental data presented in the previous section can be fitted with an existing governing model. As already stated, it is apparent from Figure 4 that the weakening is not linear, and therefore we will disregard the idealized linear slip–weakening model [Andrews, 1976] and we will focus only on non linear functions. In particular, we consider prominent examples of the two main classes of governing laws proposed so far. The first class is represented by the rate– and state–dependent friction laws. They are known to be able to reproduce a lot of features of crustal earthquakes (as well as the entire seismic cycle), but they have the obvious weakness of being unable to deal with zero sliding speed. In order to solve this issue some efforts have been made using grain–scale creep process equations [Moore et al., 2019] or other analytical manipulations [Perrin et al., 1995; Nielsen et al., 2000; see also Cocco et al., 2004]. In the present study we consider both the canonical ageing, Dieterich–Ruina [DR thereinafter; see Dieterich, 1978] formulation:

$$\begin{cases} \tau = \left[\mu_* + a\ln\left(\dfrac{v}{v_*}\right) + b\ln\left(\dfrac{\Psi v_*}{L}\right)\right]\sigma_n^{eff} \\ \dfrac{\mathrm{d}}{\mathrm{d}t}\Psi = 1 - \dfrac{\Psi v}{L} \end{cases} \quad (2)$$

and the canonical slip, Ruina–Dieterich [RD henceforth; see Ruina, 1983] law:

$$\begin{cases} \tau = \left[\mu_* + a\ln\left(\dfrac{v}{v_*}\right) + b\ln\left(\dfrac{\Psi v_*}{L}\right)\right]\sigma_n^{eff} \\ \dfrac{\mathrm{d}}{\mathrm{d}t}\Psi = -\dfrac{\Psi v}{L}\ln\left(\dfrac{\Psi v}{L}\right) \end{cases} \quad (3)$$

In equations (2) and (3) $\mu_*$ and $v_*$ are reference values for the friction coefficient and slip velocity, respectively, $a$ and $b$ are dimensionless governing parameters and $L$ is the length scale over which the single state variable $\Psi$ evolves. It is well known [e.g., Bizzarri and Cocco, 2003] that $L$ is related to the (equivalent) slip–weakening distance $d_0$ by some multiplying factor, which depends on the specific analytical expression of the rate– and state– function.

The other constitutive law considered for the fit of laboratory data is the non linear slip–dependent friction, as proposed by Ionescu and Campillo [1999] and by Voisin [2002] in their 2–D, numerical simulation of a spontaneous, mode III (anti–plane) rupture propagation:

$$\mu = \begin{cases} \mu_u - \dfrac{\mu_u - \mu_f}{d_0}\left(u - p_{IC}\sin\left(\dfrac{2\pi u}{d_0}\right)\right), & u < d_0 \\ \mu_f, & u \geq d_0 \end{cases} \quad (4)$$

where $\mu_u$ and $\mu_f$ are the peak and residual values of friction coefficient, respectively, and $p_{IC}$ is a governing parameter having the dimension of slip. Note that the friction coefficient is equal to $\mu_f$ when the slip reach $d_0$.

To fit experimental data we have adopted a Levemberg Marquardt algorithm for all the considered functional forms. In the case of the DR and RD laws things are made more complicated by the fact that the time dependence of state variable $\Psi$ in the first equations in (2) and in (3) has not an explicit form, but it must be obtained from its derivative, given by the second equations in (2) and in (3). The calculation is therefore not straightforward and it requires to perform an integration during the fitting process. Moreover, the integral not only involves time but also slip velocity. In Bizzarri and Petri [2016] it has been shown that, in the case of DR law, the variable $\Psi$ obeys a recursion formula that can be usefully employed for numerical integration:

$$\Psi(t) = \Psi(0) + e^{x(t)/L}\int_0^t e^{x(t')/L}\,\mathrm{d}t' \quad (5)$$

Discretizing time into identical intervals $\delta t$ leads to

$$\Psi_{n+1} = \Psi_n e^{-\delta x_{n+1}/L} + \delta t \quad (6)$$

**Andrea Bizzarri et al.**

where $\delta x_n = x(t_n) - x(t_{n-1}) = v_n \delta t$. This expression has been employed in a Python-based code to fit the experimental friction curves with the DR model (equations (2)). In the case of RD law it has been used a Runge–Kutta algorithm.

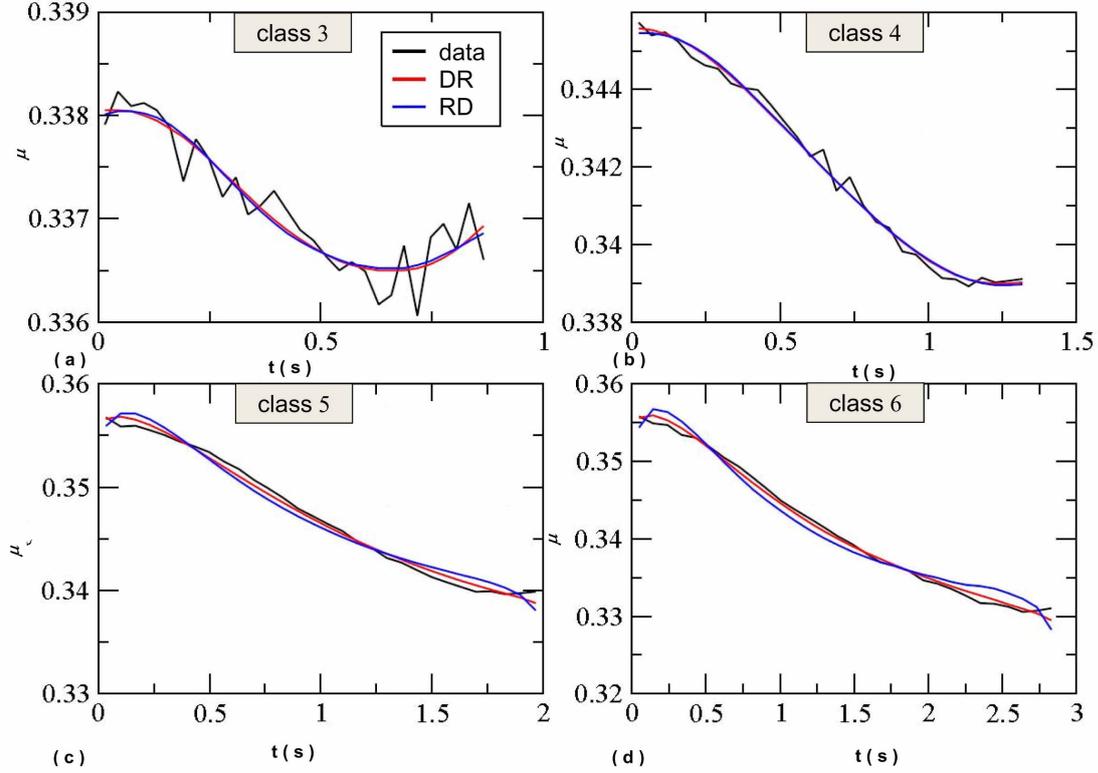

**Figure 5.** Average friction coefficient as a function of time for the four classes (4 to 6) of Figure 3. Experimental data (exp in the caption) are reported in black. Each panel also shows, as red and blue curves, the best fitting with the Dieterich–Ruina (DR) and Ruina–Dieterich (RD) governing models (equations (2) and (3), respectively). The fitting procedure is described in section 3 and the fitting parameters are reported in Table 2.

| class | model | $\mu_*$ | $a$ | $b$ | $L$ (cm) | $v_*$ (cm/s) | $\chi^2$ | $R^2$ |
|---|---|---|---|---|---|---|---|---|
| 3 | Dieterich | 0.336 | 2.346E-09 | 1.138E-03 | 5.254E-02 | 8.416E-02 | 3.98E-06 | 0.877 |
|   | Ruina | 0.333 | 5.519E-21 | 1.440E-03 | 5.978E-01 | 1.178E-01 | 4.01E-06 | 0.877 |
| 4 | Dieterich | 0.331 | 2.397E-05 | 4.206E-04 | 4.753E-02 | 2.780E-08 | 3.47E-06 | 0.992 |
|   | Ruina | 6.2E-9 | 3.019E-42 | 8.709E-02 | 9.751E+00 | 1.376E-01 | 3.86E-06 | 0.992 |
| 5 | Dieterich | 7.2E-9 | 5.102E-04 | 2.464E-02 | 1.836E-00 | 1.132E-03 | 2.09E-05 | 0.993 |
|   | Ruina | 0.271 | 1.519E-03 | 1.661E-03 | 1.470E-00 | 1.576E-08 | 6.12E-05 | 0.980 |
| 6 | Dieterich | 1.9E-8 | 9.518E-04 | 2.469E-02 | 1.779E-00 | 1.212E-03 | 2.90E-05 | 0.995 |
|   | Ruina | 0.214 | 2.061E-03 | 3.107E-03 | 3.198E-00 | 7.618E-09 | 6.98E-05 | 0.989 |

**Table 2.** Fitting parameters in the case of rate - and state - dependent friction laws (see section 3 for details). To quantify the goodness - of - fit we also report the value of $\chi^2 = \sum_{i=1}^{N} \frac{(\mu_i^{(exp)} - \mu_i^{(fit)})^2}{\mu_i^{(fit)}}$ and $R^2 = 1 - \sum_{i=1}^{N} \frac{(\mu_i^{(exp)} - \mu_i^{(fit)})^2}{(\mu_i^{(exp)} - \overline{\mu^{(exp)}})^2}$, where $N$ is the number of experimental data, superscripts exp and fit stand for laboratory data and analytical fit, respectively, and the operator – indicates the average value of the series.

The fit with all free parameters is very slowly converging and tends to produce extremely large values of $v_*$ and $L$ and a negative (unphysical) value of $\mu_*$. We have imposed the physical constrain $a > 0$, $b > 0$ and we have looked for converging cases with $b > a$ (namely, a velocity–weakening regime, which is known to identify an unstable regime; see for instance Gu et al. [1984]). The whole ensemble of the fitting parameters are listed in Table 2 and the results are plotted in Figure 5. It is apparent from Figure 5 that, although the qualitative behavior is generally captured, this friction law is not totally suitable to describe experimental results. In particular, it is clear that the DR and the RD models for class 3, where the oscillations in data are more significant, are able to reproduce the general trend. However, for classes 5 and 6 they do not capture the last stage of the friction, in that they are unable to give a clear definition of the equivalent slip–weakening distance. Moreover, the rate and state laws tend to flatten the experimental data. We also note that the result from Table 2 is that the curve are basically state–dependent, since the fit provides that the parameter $a$ is often negligible with respect to $b$ (in other words, the fit tends to suppress the dependence on the velocity).

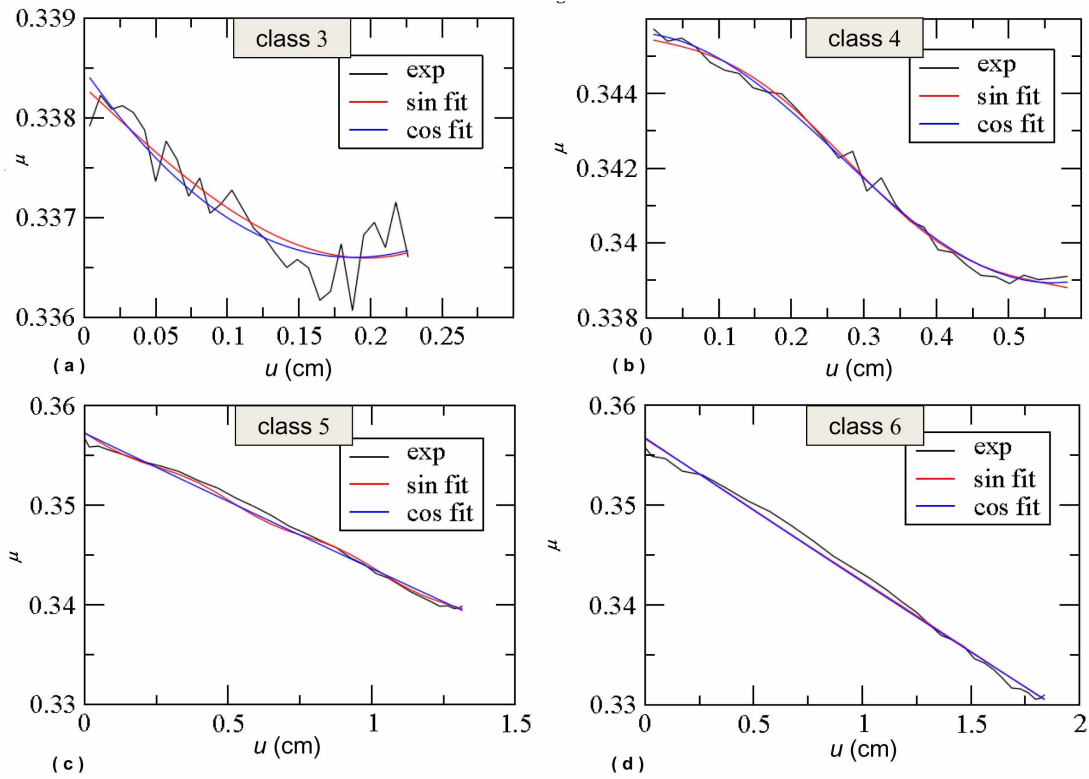

**Figure 6.** Results for the fit of the laboratory data (black curves) of Figure 3 with the Ionescu and Campillo expression (equation (4), red curves). Just for comparison we have also reported (as blue curves) the results with a cosine function. The fitting procedure is described in section 3 and the fitting parameters are reported in Table 3.

The use of the friction law with the expression of equation (4) suggested by Ionescu and Campillo [1999] yields better fits; the parameters of the fitting procedure are tabulated in Table 3 and the corresponding results are reported in red in Figure 6. Also in this case we can observe that the fitting parameters have severe problems; sometimes $d_0$ assumes extraordinarily large values, often $\mu_u$ is nearly equal to $\mu_f$. For the class 3, which has more fluctuations with respect to the other classes (Figure 6a), the analytical formulation fits only qualitatively the experimental data. For classes 4 and 6 (Figures 6b and 6d, respectively) equation (4) gives overestimated and underestimates of the laboratory data. For completeness we have also considered a fit with a cosine function. As expected, this analytical formulation (reported in blue in Figure 6) gives fits very similar to those obtained with equation (4), but sometimes with some negative parameters.

| class | function | $\mu_u$ | $\mu_f$ | $d_0$ (cm) | $p_{IC}$ (cm) | $\chi^2$ | $R^2$ |
|---|---|---|---|---|---|---|---|
| 3 | sin | 0.338 | 0.342 | 0.850 | 0.490 | 2.09E-06 | 0.923 |
|   | cos | 0.351 | 0.308 | 2.094 | -0.623 | 1.86E-06 | 0.912 |
| 4 | sin | 0.345 | 0.339 | 0.557 | 0.060 | 1.216E-06 | 0.996 |
|   | cos | 0.342 | 0.344 | 1.111 | -2.037 | 1.060E-06 | 0.997 |
| 5 | sin | 0.357 | 0.350 | 0.512 | -0.027 | 5.615E-06 | 0.997 |
|   | cos | 0.931 | -2.226 | 228.6 | -41.57 | 7.980E-06 | 0.996 |
| 6 | sin | 0.357 | 0.294 | 2.099 | 0.002 | 1.096E-05 | 0.997 |
|   | cos | 0.933 | -2.525 | 236.7 | -39.462 | 1.256E-05 | 0.997 |

**Table 3.** Fitting parameters in the case of non linear slip-weakening friction law (see section 3 for details). The function sin denotes the original Ionescu and Campillo model (equation (4)), while cos indicates the fit with the same expression, but with cosinus function instead of sinus function.

# 4. Numerical modeling

## 4.1 Methodology

We consider spontaneous (i.e., the velocity at which the rupture advances on the fault plane is not prescribed a priori, but it is a part of the solution of the problem), fully dynamic (i.e., considering inertial effects over the whole time window), truly 3–D (i.e., with both mode II and mode III coupled together) ruptures, spreading over a single, planar, vertical, strike slip fault. We numerically solve the fundamental elastodynamic equation, neglecting body forces, by using a second–order accurate, finite–difference, OpenMP–parallelized code [Bizzarri and Cocco, 2005; Bizzarri, 2009] which makes it possible to easily incorporate different fault boundary conditions.

We have already seen (see section 3) that data obtained from laboratory experiments can not be exactly represented by some analytical equation presented in literature. Therefore, instead of introducing an analytical formulation of the governing model (such as slip–weakening or some rate– and state– function), we directly insert the laboratory data as obtained from experiments (see section 2.2). In such a novel approach, real data — and not an equation interpolating them — thus constrain the traction evolution on the fault.

We assume that the experimental behavior is representative not only of a specific fault point, but of the whole fault surface. This reflects into the assumption that the rheological properties of the real fault are spatially homogeneous; this of course is a conservative assumption, because we know that the slip complexity observed during real–world earthquakes is often due to (or numerically reproduced through) the insertion of spatially inhomogeneous distributions of governing parameters.

Due to the necessity of properly solving the spatial and temporal scales of the breakdown process, we need to impose a very fine time discretization ($\Delta t \sim 1$ ms; see Table 4). Consequently, the numerical discretization of the fault slip (as resulting from the numerical algorithm) is significantly finer compared to the sampling of the laboratory data after averaging. Let ($u^{[m]}$, $\tau^{[m]}$) be the couple of (fault slip, fault traction) at the actual (updated) time level $m$ within the numerical algorithm. Since we want to have real data as fault boundary condition, then $\tau^{[m]}$ has to be constrained by laboratory data. To this goal, we interpolate laboratory data (namely, each time series (linear displacement, traction)) by using a cubic spline algorithm to have the value of fault traction exactly at the value of $u^{[m]}$. Once the friction drop is completed, we prolong the experimental time series by keeping the friction constant at the value of $\mu_f$; in this way, we guarantee that the fault strength is determined for arbitrarily large values of fault slip, even when the dynamically computed fault slip exceeds the maximum linear displacement measured during a laboratory experiment. This apparent artifact is absolutely necessary when we consider that the modeled slip of the spontaneous rupture can easily exceed that recorded in laboratory.

Since in the laboratory curves we have only the weakening state, starting from the peak stress we need some mechanism which rises the fault traction from the imposed initial value $\tau_0$ up to the yield stress, which namely identifies the onset of the rupture in the target fault point. This "nucleation" is imposed by following the procedure described in Bizzarri [2010]; this strategy guarantees a smooth transition between an early rupture propagation at a constant rupture speed and the subsequent spontaneous propagation over the fault plane. The fault is assumed to be embedded in a Poissonian elastic medium with homogeneous properties (see Table 4).

| Parameter | Value | |
|---|---|---|
| **Medium and Discretization Parameters** | | |
| Lamé constants, $\lambda = G$ | 24.3 GPa | |
| S wave velocity, $v_S$ | 3 km/s | |
| P wave velocity, $v_P$ | 5.196 km/s | |
| Mass density, $\rho$ | 2700 kg/m$^3$ | |
| Fault length, $L$ | 80 km | |
| Fault width, $W$ | 23 km | |
| Spatial grid spacing, $x_1 = x_2 = x_3 \equiv x$ | 50 m | |
| Time step, $t$ | 1.39 x 10$^{-3}$ s | |
| Courant–Friedrichs–Lewy ratio, $\omega_{CFL} = v_S \Delta t/\Delta x$ | 0.083 | |
| Depth of the hypocenter, $x_3^H$ | 7 km | |
| **Fault Constitutive Parameters** | | |
| | **Model A** | **Model B** |
| Experimental dataset | class 4 | class 6 |
| Magnitude of the imposed initial shear stress, $\tau_0$ | 62.198 MPa | 62.888 MPa |
| Magnitude of the assumed effective normal stress, $\sigma_n^{eff}$ | 182 MPa | |
| Static friction coefficient from laboratory data, $u_u$ | 0.34590 ($\leftrightarrow \tau_u$ = 62.95 MPa) | 0.35679 ($\leftrightarrow \tau_u$ = 64.94 MPa) |
| Dynamic friction coefficient from laboratory data, $u_f$ | 0.33898 ($\leftrightarrow \tau_f$ = 61.69 MPa) | 0.33148 ($\leftrightarrow \tau_f$ = 60.33 MPa) |
| Dynamic stress drop, $\Delta \tau_d = \tau_0 - \tau_f$ | 1.26 MPa | 4.60 MPa |
| Inferred strength parameter, $S = (\tau_u - \tau_0)/(\tau_0 - \tau_f)$ | 1.5 | 0.8 |
| Inferred (equivalent) slip–weakening distance, $d_0$ | 0.097 m | 0.31 m |

**Table 4.** Parameters used in the spontaneous 3–D models. We consider spatially homogeneous properties and a constant effective normal stress.

### 4.2 Scaling of the laboratory data

The numerical procedure described so far is fully general and it can be easily exploited for any kind of experimental data, not necessarily on granular media or friction on bare surfaces. The analytical fit of real data (such as described in section 3) is not necessary within our elastodynamic code, since we can impose, as we have seen in section 4.1, the fault boundary condition simply by inserting the data, directly as they emerge from laboratory experiments.

Of course, we are aware of the fact that for any kind of laboratory experiments we operate an extrapolation, in that we assume that the behavior of the friction curve constrained by experimental data is not changing by assuming different boundary conditions (i.e., normal stresses). This is a problem holding for any kind of experimental machine, that is, as already highlighted, inherently unable to fulfill the crustal conditions in terms of both sliding speed and confining stress.

We have seen that the experiments performed with granular media are conducted at extremely low confining stress (~ hundred of Pa), while crustal faults are expected to accommodate higher effective normal stress (~ hundred of MPa). In the simulations presented and discussed in the following of the present study we assume $\sigma_n^{eff}$ = 182 MPa, which leads to typical value of fault traction. As already stated above, the value of $\sigma_n^{eff}$ is kept constant through the whole numerical simulations (in other words, there is no pore fluid migration in our present models), so that $\sigma_n^{eff}$ represents the stress of tectonic origin only.

Moreover, we also understand that the friction drop predicted by the actual laboratory experiments is small compared to that obtained during high speed velocity on sliding surfaces, where the drop is expected to be nearly complete. First of all, we note that it has been observed that the stress drop values measured in high–speed laboratory experiments greatly exceed those commonly reported for natural earthquakes [Zielke, 2017], where the reduction of the values after the friction drop are typically not larger than 10 MPa [Scholz, 1996; Allmann, 2009] and thus comparable in relative size to what observed here. Moreover, we can interpret the friction curves obtained by granular media as proxies of natural slip events without thermal pressurization, flash heating, melting lubrication and without all the mechanisms that tend to increase the degree of instability of the fault surface and that finally tend to produce dramatically large stress releases [e.g., Bizzarri, 2014].

Another problem intimately related to the adoption of laboratory data to model real–world ruptures is represented by the value of the characteristic spatial length. As a general rule, we expect that the (equivalent) slip–weakening distance $d_0$ would change by moving from results obtained within a laboratory meter–sized sample to a crustal fault which is several kilometer long [Marone and Kilgore, 1993; see also Ohnaka, 2003]. Indeed, from our experiments on granular media we know that $d_0$ ~ 1 cm or less. In our numerical experiments, to have a parameter set compatible with estimates (or inferences) of $d_0$ for real earthquakes (~ some fractions of meter) we rescale the values of experimental linear displacement by a factor of 20.

## 5. Results of synthetic earthquakes

Among the ensemble of numerical simulations that we generate, we select two rather different cases (summarized in Table 5), which are representative of the various scenarios. Model A assumes class 4 of laboratory data (red curves in Figures 3 and 4), while Model B assumes class 6 (blue curves in Figures 3 and 4). The value of the initial shear stress $\tau_0$ causes Model A be subshear, while Model B supershear. The general conclusions discussed below do not change by selecting other classes of laboratory data and other values of initial shear stress.

Figures 7 and 9 show the snapshots of fault slip velocity and fault traction at the same time level. As expected, the supershear event (Model B) has enlarged more compared to the subshear one (Model A). Moreover, it displays the typical behavior observed in dynamic ruptures controlled by canonical slip–dependent or rate– and state–dependent friction laws:

1) the elongation in the direction of prevailing mode II (i.e., along the strike direction). This greater elongation of the broken area culminates, in the case of supershear rupture (Model B; Figure 9), in the birth of a secondary rupture front, which accelerates up the terminal velocity of *P* wave, as theoretically expected [see for instance Bizzarri and Das, 2012 and references cited therein];
2) the increase of fault slip velocity in the correspondence of the rupture front (the separation between the unbroken part of the fault and the region which is already slipping), especially in the region where the mode II is dominating with respect to the mode III (Figures 7a and 9a);
3) the stress concentration near the rupture tip (Figures 9a and 9b), which is responsible of the crack advance and of the shrinking of the cohesive zone [e.g., Andrews, 1976];
4) due to the more significant stress drop (4.60 MPa for Model B against 1.26 MPa for Model A), also the peak fault slip velocity is greater in the supershear model compared to the subshear one (compare Figures 7a and 9a; see also Figures 8c and 10c).

In Figures 8 and 10 we present the result pertaining to one on–fault receiver (identified by the red triangle in Figures 7 and 9). We can see that the imposed fault boundary condition (i.e., the data from laboratory experiment) is perfectly reproduced by the numerical model (the superimposition of pink and black curves in Figures 8b and 10b indicates a perfect matching). The small decrease of traction just before the peak (visible in Figures 8a and 10a)

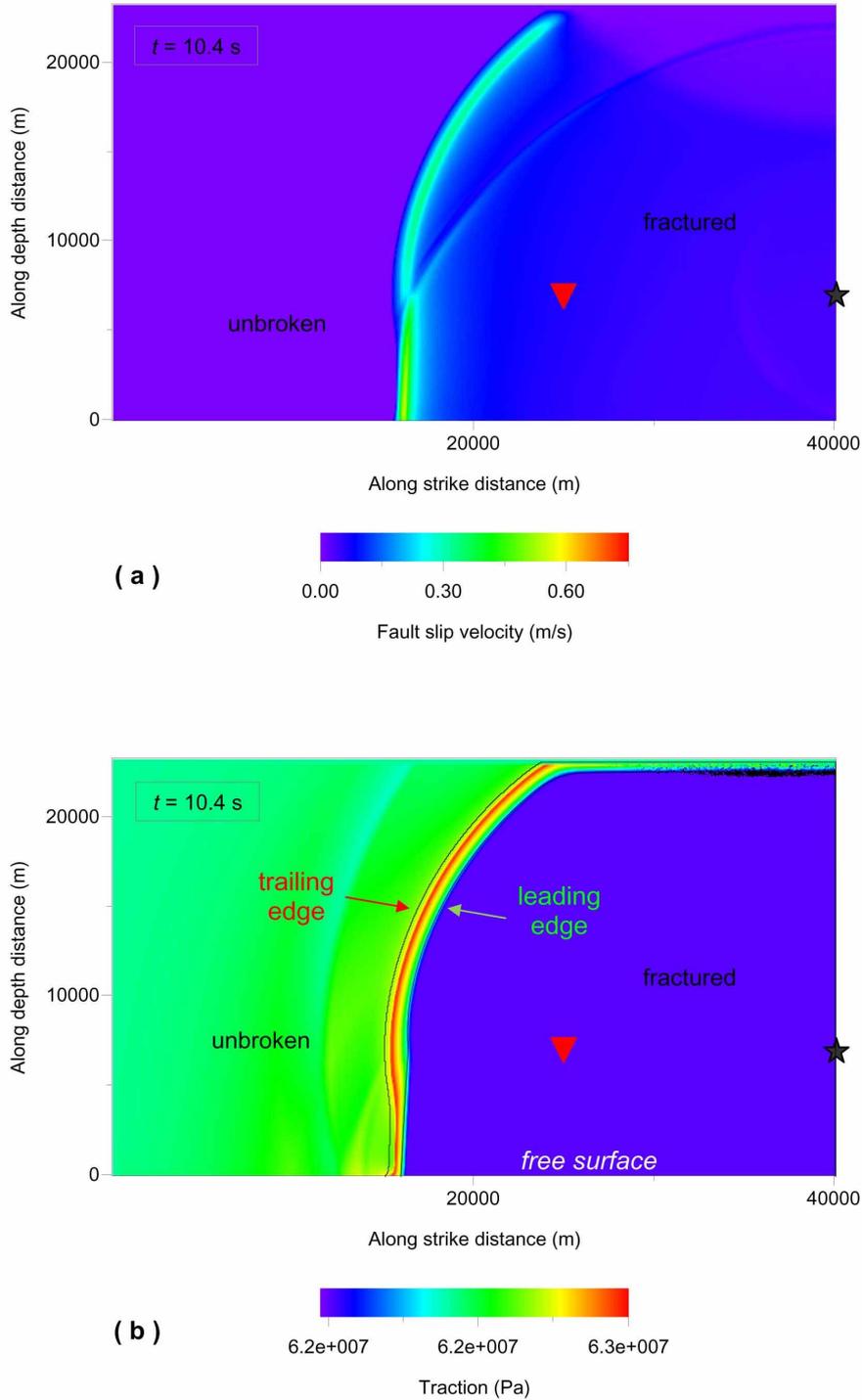

**Figure 7.** Results of spontaneous rupture propagation in a 3–D fault governed by friction behavior as derived in laboratory (class 4; Model A). Snapshot of fault slip velocity (panel (a)) and traction (panel (b)) at time $t$ = 10.4 s. In panel (b) is also indicated, as continuous black lines, the extension of the so–called cohesive zone, defined as the spatial region where the traction is in between the upper level and the final value. For sake of simplicity, due to the symmetry of the problem, only a half of the fault surface is reported in the plots. Except for the free surface (where the free–of–traction condition is imposed), the other terminal faces of the computational domain are absorbing boundaries, in order to avoid spurious back–reflections of signals into the model. (Note that the free surface is responsible of some reflections, which are physical and not numerical noise.) The black star indicates the location of imposed hypocenter. The numerical approach is described in section 4, while parameters are listed in Table 4 (Model A).

**Andrea Bizzarri et al.**

is not a numerical pollution, but it is the interference with the front emerging when the rupture tip hits the free surface and therefore a signal is back–propagated into the model. As theoretically expected, the breakdown zone time $T_b$ (namely, the time requested for the stress drop to be completed) is lower in the supershear case compared to the subshear one; from Figures 8a and 10a it emerges that the drop requires 0.67 s in the case of Model A and 0.40 s in the case of Model B. As stated above, this is not surprising, because it is well known that supershear ruptures exhibit a more abrupt stress release compared to subshear ones [see also Appendix A of Bizzarri et al., 2001].

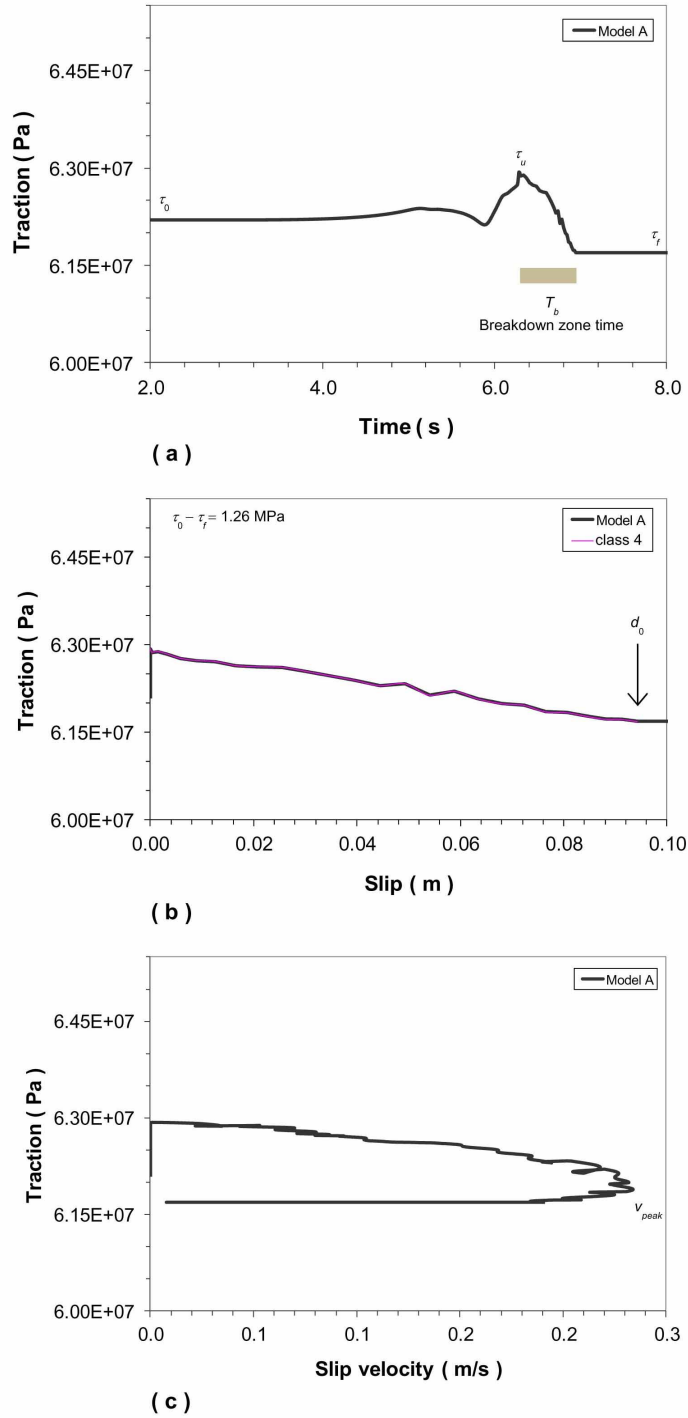

**Figure 8.** On–fault solutions pertaining to Model A. The receiver is located at the hypocentral depth (7 km) and at a distance of 15 km from the imposed hypocenter (red triangle in Figure 7); at this location the rupture is well developed and it is totally spontaneous. (a) Time evolution of fault traction. (b) Slip–weakening curve (i.e., traction as a function of cumulated fault slip). The experimental curve (namely, the boundary condition represented by the original laboratory data, imposed as governing model, is superimposed to the solution of the numerical code. (c) Phase portrait (i.e, traction as a function of fault slip velocity).

It is well known that the breakdown zone time does not depend only on $\tau_u$, $\tau_f$ and $d_0$ (that we impose as fault boundary conditions as emerging from the laboratory records), but also on the value of the initial stress $\tau_0$ (that we impose as input parameter to simulate sub– and supershear ruptures). Moreover, $T_b$, which is a result of the model and not a assigned parameter such as $d_0$, also depends on the collective behavior of our extended 3–D fault.

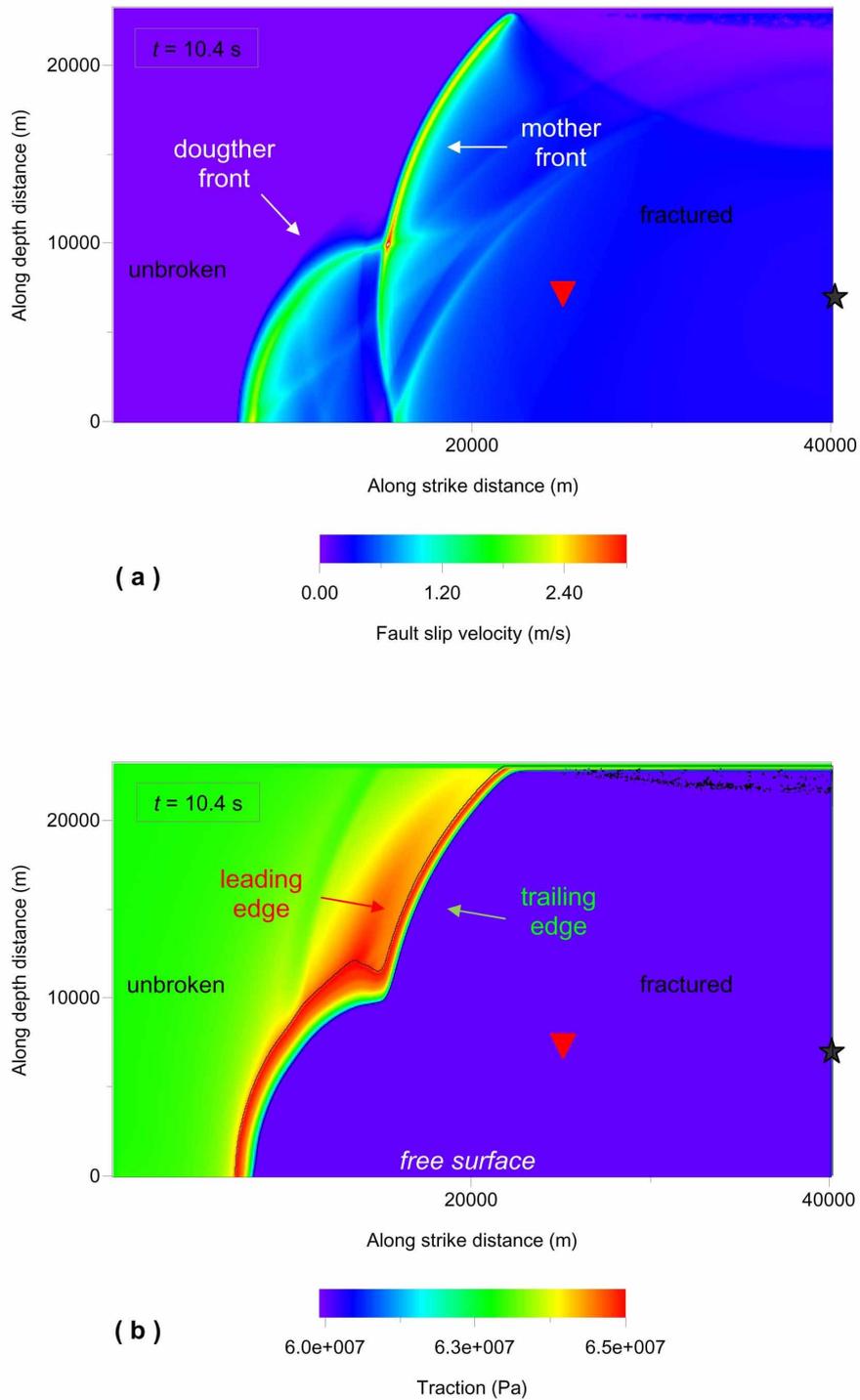

**Figure 9.** The same as Figure 7, but now for Model B (see Table 4).

**Andrea Bizzarri et al.**

In this perspective it is not surprising that the resulting $T_b$ does not necessarily correspond to that measured in the laboratory experiments (specifically, for class 4 it is 1.13 s and for class 6 it is 2.6 s; see Figure 3c).

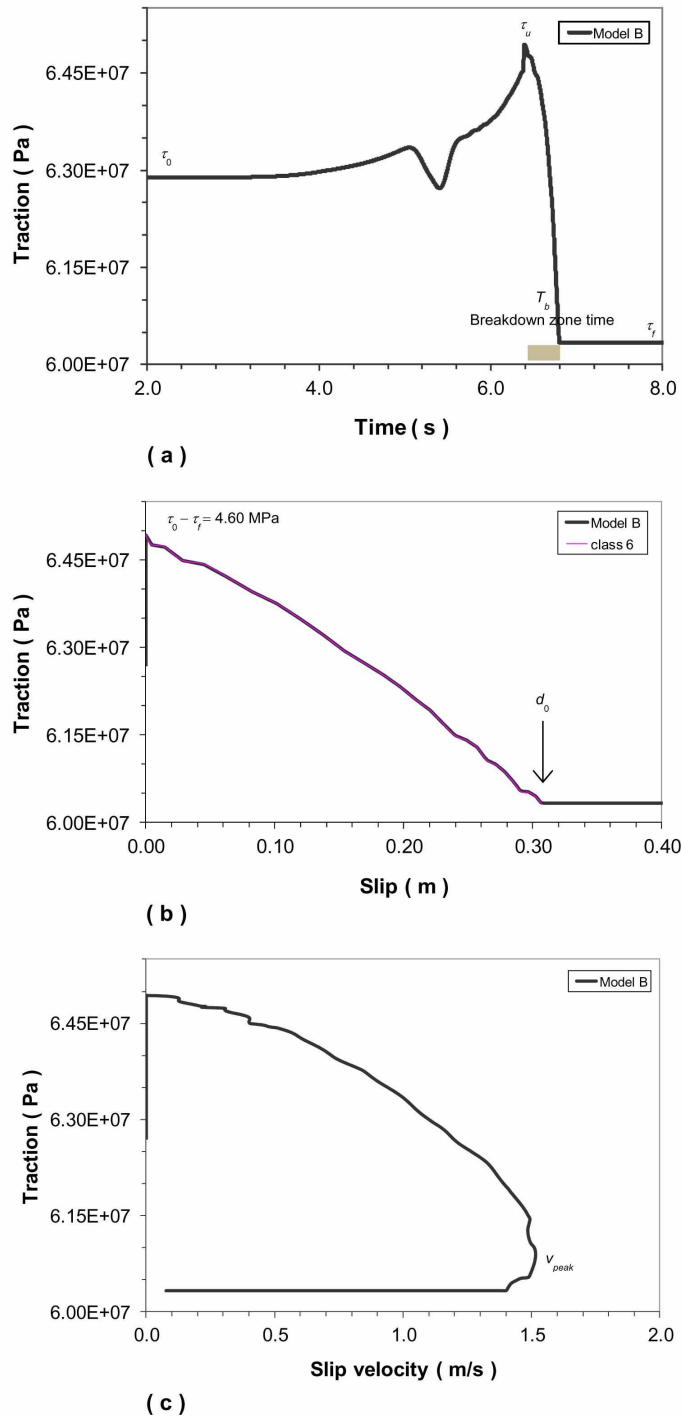

**Figure 10.** The same as Figure 8, but now for Model B (see Table 4).

The same holds for the fault slip velocity. In our procedure we do not impose the time history of $v$ (we have a spontaneous model and not a kinematic model where the source time function is assigned a priori to retrieve the traction history). The time evolution of the fault slip velocity depends again on the degree of the instability of the fault (namely, on the value of the strength parameter *S*), but also on the presence of the free surface, that interacts with the rupture front (the fault is not buried and can hit the free surface), and on the coexistence of the mode II and mode III of fracture (natural consequence of the 3–D nature of the model). Moreover, the conservative assumption of constant final value of the frictional resistance (recall that $\tau = \tau_f$ once the breakdown process is completed and no fast restrengthening is incorporated into the model) prevents healing phenomena and thus avoids that *v* turns back to null values. Indeed, the phase diagrams shown in Figures 8c and 10c are consistent with models using other friction laws [see, just for an example, Figures 3a and 2a of Tinti et al., 2004].

## 6. Discussion and conclusive remarks

In this paper we have presented the rheological behavior emerging from laboratory shear experiments, performed on granular materials [Baldassarri et al., 2006, 2019; Petri et al., 2008].

The role of granular gouge in slip and rupture propagation is a continuously investigated and debated issue, not only in seismology [Mingadi, 2021]. The difficulty, or impossibility, to access real structures seems in any case to prevent detailed knowledge about the dynamics of specific faults. Nevertheless, the similarities observed in different phenomena driven by shear stress, prospects the possibility that many different situations could result in macroscopic behaviors classifiable into a not too large number of typical behaviors, at least for some aspects. As an example, a recent study [Denisov, 2017] shows that, despite made of elementary constituents very different in size and interaction, a granular medium and a metallic glass display strong similarities in their shear dynamics and weakening inertial effects. This property, often called universality, is related to what happens in other physical phenomena, specifically the critical ones [Sethna, 2001; Petri, 2019].

It has also been shown in laboratory faults that the presence of granular gauge makes easy for seismic waves to trigger weak or critical faults [Johnson, 2005]. To this respect it is worth to mention that the apparent friction coefficient for matter in the form of grains spans a wide range of values but is generally much lower than the one of the constituent solid material. This feature, which can have dramatic effects on other phenomena like avalanches and landslides, has been observed in very different conditions and materials; millimetric glass beads at a pressure of 1 kPa and sheared at some mm/s [Baldassarri, 2019], microsized particles of at tenths of MPa sheared at few $\mu$m/s [Mair, 2009], generic experiments [GDR MiDi, 2004]. Low values of the apparent friction coefficient have been also found in numerical simulations performed at pressures reaching 70 MPa, with load point velocity ranging from 1 micron/s to 10 mm/s) [Mair, 2007; other examples are in Tang et al., 2019 and references therein]. The presence of weak faults with friction as low as $0.1 < \mu < 0.3$, and its consequences on seismic activity, has been recently analyzed and discussed in [Collettini, 2019].

It is well–known [e.g., Ohnaka, 2003] that real earthquakes can be regarded as a mixture of shear friction on pre–existing surfaces and fracture of intact rocks (i.e., the formation of new discontinuity interfaces). The granular flow experiments considered here insert another ingredient into the attempt to understand the rupture propagation processes.

The experimental data provide a behavior of the sliding velocity (Figure 3b) which is compatible with theoretical expectation and numerical modeling of ruptures. Indeed, the measured velocity closely resembles the so–called modified Yoffe function [Tinti et al., 2005; see also Broberg, 1978, 1999; Freund, 1979], which is a modification of the singular function describing the steady state solution of a mode I rupture. Moreover, the behavior of the experimental velocity is also compatible with the closed–form analytical solution that Bizzarri [2012a] found by considering a 1–D spring–slider analogue fault system subject to the linear slip–weakening friction law.

We have shown that this behavior cannot be interpreted in the framework of the rate– and state–dependent friction laws; indeed Figure 5 clearly indicate that data are not well described by the Dieterich–Ruina nor Ruina–Dieterich governing models [equation (2); Dieterich, 1978; Ruina, 1983]. On the other hand, we have also seen that a sinusoidal slip–dependent friction law, such as that proposed by Ionescu and Campillo [1999] and Voisin [2002] (equation (4)), is not able to capture all the features of the experimental curve (Figure 6). Although the fit seems to be of great quality merely from a numerical point of view ($R^2$ is always greater than 0.85) the fit produces values that are sometime controversial (see Tables 2 and 4 and section 3) and often suppresses the dependence on the slip velocity, thus giving only a state–dependent friction.

**Andrea Bizzarri et al.**

The considered models are quite representative of two rather different classes of constitutive laws, the slip–dependent and the rate and state friction. Of course, in the literature different models have been proposed, but since the observed stress reduction on our granular experiments is not dramatic, we do not consider other governing equations proposed in the recent literature, which account for large friction reduction at high speed [see for instance Nielsen et al., 2021]. It is important to mention that in a separate paper we will try to construct a mathematical model able to describe the data arising from experiments on granular materials, possibly resorting to microphysical models [Van der Ende et al., 2018].

We want to emphasize that the fitting procedure presented here (see section 3) is totally novel and it represents the first attempt to fit the friction value of a rate and state function to spontaneous slip; the only other attempt presented in literature, to our knowledge, is the work by Birthe [2015], based on a solid–on–solid experiment, where the parameters are estimated by fitting the velocity, and the 22 % of the considered data could not be fitted satisfyingly with the Dieterich–Ruina model. Indeed, difficulties in fitting data have already been found from both experiments [see, e. g., Scuderi, 2017], microphysical models [Van der Ende et al., 2018] and granular models [Ferdowsi and Rubin, 2020], who also reported that none of the existing rate and state laws reproduce all the robust features emerging from laboratory data.

In the second part of the present paper we exploit a novel numerical approach which incorporates the data directly as boundary condition within the numerical modeling of rupture propagation (section 4). In particular, we have directly inserted the laboratory time series in a numerical code which solves the elastodynamic fundamental equation for a truly 3–D problem on a single, planar, vertical, strike slip fault [Bizzarri and Cocco, 2005; Bizzarri, 2009]. We have assumed that the shape of the friction curve obtained by the experimental data is not changing moving from the scale of laboratory to that of real–world faults and we also neglect fluids migrations and other additional weakening mechanisms. Indeed, there is no reason to overemphasize here that this extrapolation is performed in all cases in which modelers use a laboratory–derived friction model (rate– and state–dependent laws, slip–dependent laws, etc.) to simulate the dynamics of crustal earthquakes. In order to model the physics of earthquakes developing on a crustal fault 80–km long, we simply change the effective normal stress acting on the slip zone (which of course is several order of magnitude greater than that imposed in the granular material experiments) and the scaling distance over which the traction evolves, in agreement with usual practice [Marone and Kilgore, 1993]. Moreover, we also assume that the characteristic length is constant (uniform) over the whole extended fault plane (in other words, the punctual value retrieved from the laboratory data is assumed to be representative — after the adequate scaling — of the whole fault surface. This assumption is conservative, although we know that the weakening distance can be variable [see, for instance, Mair and Marone, 1999 and Leeman et al., 2018]; this issue has been considered in the theoretical paper by Bizzarri [2021].

Moreover, since the fracture experiments of Ohnaka and Yamashita [1989] we know that the friction is not a single–valued function of slip velocity (see their Figure 2b); it is therefore natural to consider the friction dependent on the slip.

Our goal is not to reproduce, in the same scale, with a numerical procedure the results obtained in the laboratory, but is to explore whether the frictional behavior obtained in laboratory experiments on granular material is dynamically consistent, i.e., if it is able to describe the main features expected in the spontaneous propagation of a rupture in an extended fault having realistic dimension and initial conditions. At the same time, we understand that the rheology coming from lab experiments namely pertain to a viscous behavior, while we numerically model a slip failure as propagating shear rupture on a fault plane. Despite to this conceptual difference, our numerical results indicate that the rheological behavior emerging from shear experiments on granular materials is dynamically consistent, reinforcing the similarities between fracture and granular flow [see, e.g. Herrera et al., 2011]. Moreover, the discrete element model used in Van den Ende and Niemeijer [2018] suggests that granular materials frictional behavior is suited to model stick–slip phenomenon. In other words, the numerical experiments presented here indicate that granular materials can be able to describe the spontaneous rupture propagation of a crustal earthquake.

Indeed, we can model the propagation of a truly 3–D rupture (which is a mixture of shear modes II and III), spontaneously accelerating up to some terminal rupture speed. The numerical models exhibit all the relevant features of dynamic rupture propagation, i.e., the pseudo–elliptical shape of the rupture front (with an elongation in the direction of prevailing mode II), the stress concentration and the maximum fault slip velocity near the rupture tip. Moreover, with such rheological curves, it is possible to reproduce both sub– and supershear events, i.e., earthquakes traveling below or above the *S* wave speed of the elastic medium in which the fault structure is embedded.

We also emphasize that the data obtained from laboratory experiments of granular material, although conceptually different from those emerging from fracture experiments on intact rocks and from canonical frictional experiments, produce results of dynamic propagation of slip failure totally consistent.

At a more general level, this work is intended to better link together the laboratory and the numerical modeling of unstable slip processes. The novel approach proposed in this study (section 4) can be exploited for any kind of experimental curve obtained with various laboratory apparatus (fracture, friction, rotary friction, etc.). The approach proposed here is not simply to understand laboratory experiments, but to directly inject the laboratory data into a numerical model. We are available to share our numerical procedure and data with other experimentalists in order to test their data and see whether they are dynamically consistent, and to see their ability to explain and understand the mechanical processes observed in nature, and potentially acting in a competing way, during a failure episode or a seismic sequence.

**Acknowledgements.** The laboratory data used in the present paper comes from laboratory experiments performed using the apparatus presented in the cited references of Baldassarri et al. [2006, 2019] and Petri et al. [2008]. We would thank the Associate Editor, Gaetano Festa, and the two anonymous reviewers who contributed to improve the manuscript.

*CORRESPONDING AUTHOR: Andrea BIZZARRI,
Istituto Nazionale di Geofisica e Vulcanologia,
Via Donato Creti 12, Bologna, Italy;
e-mail: andrea.bizzari@ingv.it